\documentclass[conference]{IEEEtran}
\IEEEoverridecommandlockouts
\usepackage{cite}
\usepackage{amsmath,amssymb,amsfonts}
\usepackage{algorithmic}
\usepackage{graphicx}
\usepackage{textcomp}
\usepackage{xcolor}
\usepackage{soul}
\def\BibTeX{{\rm B\kern-.05em{\sc i\kern-.025em b}\kern-.08em
    T\kern-.1667em\lower.7ex\hbox{E}\kern-.125emX}}
\begin{document}

\title{ML-Enabled Eavesdropper Detection in Beyond 5G IIoT Networks}

\author{
    \IEEEauthorblockN{Maria-Lamprini A. Bartsioka\IEEEauthorrefmark{1}, Ioannis A. Bartsiokas\IEEEauthorrefmark{1}, Panagiotis K. Gkonis\IEEEauthorrefmark{2}, \\Dimitra I. Kaklamani\IEEEauthorrefmark{1} and Iakovos S. Venieris\IEEEauthorrefmark{3}}
    \IEEEauthorblockA{\IEEEauthorrefmark{1}\textit{Microwave and Fiber Optics Laboratory, School of Electrical and Computer Engineering,} \\\textit{National Technical University of Athens, 9 Heroon Polytechneiou str, Zografou 15780, Athens, Greece}}
    \IEEEauthorblockA{\IEEEauthorrefmark{2}\textit{Department of Digital Industry Technologies, National and Kapodistrian University of Athens, Sterea Ellada,} \\ \textit{34400 Dirfies Messapies, Greece}}
    \IEEEauthorblockA{\IEEEauthorrefmark{3}\textit{Intelligent Communications and Broadband Networks Laboratory, School of Electrical and Computer Engineering,} \\ \textit{National Technical University of Athens, 9 Heroon Polytechneiou str, Zografou 15780, Athens, Greece}}
Emails: bartsiokamarilina@mail.ntua.gr, giannismpartsiokas@mail.ntua.gr, pgkonis@dind.uoa.gr, \\ dkaklam@mail.ntua.gr, venieris@cs.ntua.gr}

\maketitle

\begin{abstract}
Advanced fifth generation (5G) and beyond (B5G) communication networks have revolutionized wireless technologies, supporting ultra-high data rates, low latency, and massive connectivity. However, they also introduce vulnerabilities, particularly in decentralized Industrial Internet of Things (IIoT) environments. Traditional cryptographic methods struggle with scalability and complexity, leading researchers to explore Artificial Intelligence (AI)-driven physical layer techniques for secure communications. In this context, this paper focuses on the utilization of Machine and Deep Learning (ML/DL) techniques to tackle with the common problem of eavesdropping detection. To this end, a simulated industrial B5G heterogeneous wireless network is used to evaluate the performance of various ML/DL models, including Random Forests (RF), Deep Convolutional Neural Networks (DCNN), and Long Short-Term Memory (LSTM) networks. These models classify users as either legitimate or malicious ones based on channel state information (CSI), position data, and transmission power. According to the presented numerical results, DCNN and RF models achieve a detection accuracy approaching 100$\backslash$\%  in identifying eavesdroppers with zero false alarms. In general, this work underlines the great potential of combining AI and Physical Layer Security (PLS) for next-generation wireless networks in order to address evolving security threats.
 
\end{abstract}


\section{Introduction}

Fifth-generation (5G) wireless networks have been standardized and deployed worldwide, providing multimedia services at unprecedented speed and reduced cost. In general, there are two main areas of 5G use cases. The first pertains to enhanced mobile broadband services, aimed at delivering optimal quality of service (QoS) and experience (QoE) levels for applications such as ultra-high-definition (UHD) video, augmented reality (AR) and virtual reality (VR). The second category focuses on the Internet of Things (IoT), which enables ultra-large capacity and supports a massive number of connected devices. Therefore, the full-scale deployment of 5G networks can leverage enhanced mobile broadband (eMBB), massive machine-type communications (mMTC), and ultra-reliable low-latency communications (uRLLC) \cite{b1}. These requirements have been fundamental in shaping the development of a new 5G system concept based on open-source principles, facilitating the creation of new ecosystems involving all stakeholders and decentralizing data centers to edge locations closer to end-users \cite{b2}. The deployment of 5G networks is underpinned by innovative technologies such as cloud computing, massive multiple-input multiple-output (mMIMO) antenna orientations, and millimeter-wave (mmWave) transmission \cite{b3}.

While continuous connectivity and high-speed data transmission have significantly advanced communication capabilities, the introduction of 5G, along with the establishment of the New Radio (NR) physical layer, has increased the susceptibility of communication systems to security and privacy threats. The growing number of low-cost interconnected user equipment (UE) and access points has led to a noticeable increase in wireless network data traffic, accompanied by greater diversity in communication services and devices. The number of Internet-connected devices is projected to reach 40 billion by 2033, according to Statista, generating an estimated 181 zettabytes of data annually by 2025, reflecting a 23\% rise compared to 2024 figures \cite{b4}, \cite{b5}. Furthermore, decentralized 5G applications enable UEs to join and exit the network dynamically, thus introducing additional vulnerabilities in securing data transmission. Initial efforts by network designers to mitigate these risks have focused on cryptographic authentication. However, these techniques phase certain limitations, such as the need for individual secret keys for every communication link, additive latency and computational overhead, and the risk that powerful adversaries could resolve complex mathematical problems to bypass encryption \cite{b6}.

In this evolving landscape, physical layer security (PLS) has emerged as a critical domain, leveraging the inherent randomness of wireless channels to ensure confidentiality and integrity from the initial stages of transmission. Additionally, physical layer authentication (PLA) offers the capability to instantly verify legitimate UEs before signal demodulation and decoding, thereby reducing unnecessary signal processing for illegitimate transmissions. PLS techniques can also serve as an additional security layer, complementing existing measures to provide robust protection for beyond 5G (B5G) networks. To this end, various PLS strategies have been developed, including artificial noise injection to degrade eavesdroppers' signal quality, secure beamforming to block signal reception by eavesdroppers, and automatic modulation classification for intrusion detection \cite{b5}. However, they remain vulnerable to physical layer attacks introduced by new wireless technologies in the B5G era. For instance, mMIMO systems require accurate channel state information (CSI) for effective beamforming, and non-orthogonal multiple access (NOMA) is vulnerable to pilot contamination and eavesdropping attacks \cite{b5}, \cite{b7}. Consequently, researchers are increasingly exploring advanced PLS techniques enhanced by the capabilities of machine and deep learning (ML/DL). These algorithms can learn the normal and abnormal behavior of wireless networks, based on communication patterns between UEs and base stations (BS). Moreover, DL techniques can predict potential new attacks by identifying patterns from historical data, as many attacks are variations of previous ones. Currently, ML and DL are employed in various PLS applications, including automatic configuration classification, channel estimation, secure channel coding, physical layer authentication and the detection of eavesdropping attacks. The latter will be examined in depth, as it constitutes the primary focus of this paper \cite{b7}, \cite{b8}.

The rest of this manuscript is organized as follows: In Section II various previous research that has been done in this field is presented. Following is Section III, where the system model of an heterogeneous ad-hoc industrial B5G environment with the existence of legitimate users and eavesdroppers is formulated. In Section IV, different types of ML and DL algorithms for PLS are presented, while in Section V the performance evaluation as well as the performance comparison among all algorithms are described. Finally, concluding remarks are provided in Section VI. 

\section{Related Surveys}

Early efforts in the PLS-aided 5G systems are presented in \cite{b9}, where the authors propose a system designed to improve secure 5G communication by using mMIMO beamforming and security codes. A BS directs highly focused beams toward a UE in order to ensure the secure transmission of confidential data, while eavesdroppers receive considerably weaker signals. The system incorporates an additional layer of security through a specialized wiretap channel code that introduces high error rates for eavesdroppers. Simulation results with an implementable security code forced the eavesdropper to have about 0.5 bit error rate (BER). However, such a system encounters challenges, such as the need for precise channel estimation in high-dimensional matrices and the need for long secure codes that contradict with 5G low latency requirements. In this framework, authors in \cite{b10} evaluate two primary categories of secrecy performance metrics in real-world IoT scenarios. Specifically, they advocate the use of secrecy throughput as a key ergodic-based metric, which measures the average transmission rate under secrecy constraints, and they determine the optimal trade-off between transmission performance and secrecy for short-packet communication.

In more contemporary literature, authors in \cite{b11} propose various DL methodologies for network intrusion detection, focusing on Autoencoder and Deep Belief Network (DBN) approaches. Autoencoders are employed to compress input data into a feature space for outlier and intrusion detection, demonstrating high accuracy, ranging from 78\% to 98\%, especially with unsupervised pre-training. The findings indicate that DBNs constitute the most effective approach for intrusion detection due to their unsupervised learning capabilities and the ability to handle large raw data. The accuracy they achieve exceeds 93\% on all test datasets. Similarly, the authors of \cite{b12} propose three DL models to authenticate sensor nodes in an industrial wireless network scenario using CSI. Simulations and real-world experiments reveal that DNN offers the highest authentication rate, almost 1, albeit with extended training times, while Convolutional pre-processing Neural Networks (CPNN) strike a balance between performance and training time. According to \cite{b6}, ML algorithms can also be utilized for such purposes. The model utilizes the minimum time delay and maximum power of multipath components to leverage the physical layer’s channel impulse response (CIR) and distinguish legitimate users from attackers. For indoor hotspot scenario XGBoost and LGBM achieve 100\% classification accuracy, while for urban macro-cells they surpass 99\%. Lastly, \cite{b13} introduces a novel multi-module fusion neural network (MMFN) for automatic modulation classification (AMC) in pixel-coloring constellation images. This task aims to identify the modulation format of the received signal, involving a two-stage process: feature extraction and classification. Extensive simulations validate the efficacy of the MMFN-based AMC in identifying physical layer attacks in IoT networks, with classification accuracy over 80\% even at low signal-to-noise-ratios (SNRs).

As demonstrated above, the integration of advanced PLS techniques in B5G ad-hoc networks has the potential to significantly enhance security performance metrics. However, the standardization of PLS mechanisms, particularly in terms of key generation, encryption algorithms, and channel characteristics, remains an open challenge. The main contribution of our study is the comparative analysis of ML and DL models applied to B5G physical layer security, through the implementation of CSI-based authentication. This analysis was conducted using a comprehensive simulation framework that incorporates a dynamic industrial network scenario, to evaluate their effectiveness in realistic B5G environments.

\section{System Model}

\subsection{System Overview}\label{AA}
Figure 1 depicts the considered cooperative B5G ad-hoc network in an indoor factory environment. The system represents a dense industrial setup, characterized by obstacles such as walls and machinery that impact radio wave propagation. 

\begin{figure}[htbp]
\centerline{\includegraphics[width=1\linewidth]{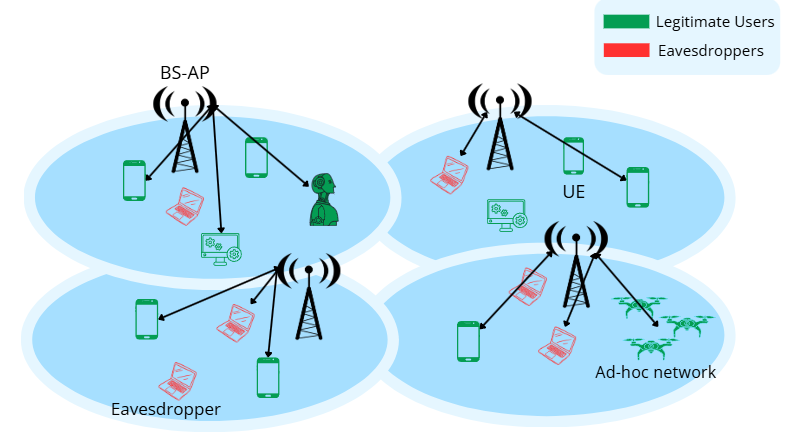}}
\caption{IIoT network topology}
\label{fig1}
\end{figure}

The scenario includes both stationary and mobile nodes, as well as ad-hoc sub-networks. The system is composed of UEs and BSs that act as Access Points (APs). There are \(N_{BS}\) BSs equipped with antenna arrays at specific angles in order to maximize coverage and reduce interference. The UEs are separated in \(N_{LU}\) typical users and \(N_{E}\) malicious eavesdroppers, each randomly distributed within the simulated environment. Both types of users can move freely and communicate in a distributed manner, without relying on any fixed infrastructure. Each UE can serve both as a transmitter and as a receiver, while BSs perform signal transmission and broadcasting. 

The system can be represented by three primary sets of entities: 
\begin{itemize}
    \item \(S_{BS} = \{b_1, b_2, …, b_{N_{BS}}\}\), the set of BSs  
    \item \(S_{LU} = \{l_1, l_2, …, l_{N_{LU}}\}\), the set of legitimate UEs 
    \item \(S_E = \{e_1, e_2, …, e_{N_E}\}\),  the set of eavesdroppers
\end{itemize}

The communication links are denoted as: \(L_{b,u}\) for links between BS-UE, and \(L_{b,e}\) for links between a BS-eavesdropper.

Delving into the placement problem, the primary goal is to position the eavesdroppers aiming to maximize the potential for signal interception while maintaining proximity to BSs and legitimate UEs. As far as legitimate UEs are concerned, their positions are distributed randomly across the simulation area. Let \({Pos}_{UE_i} = (x_{l_i}, y_{l_i}, z_{l_i})\) represent the 3D coordinates of the (i)-typical UE. Eavesdroppers are strategically placed close to BSs and legitimate UEs to simulate their ability to intercept communications. The algorithm selects a random UE, let that be \(e\) in \(S_E\). Afterward, the Euclidean distance between the selected UE and the nearest BS is calculated, where \({Pos}_{BS} = (x_b, y_b, z_b)\), as follows:
\[d(UE, BS) = \sqrt{(x_l - x_b)^2 + (y_l - y_b)^2 + (z_l - z_b)^2} \tag{1}\]
The eavesdropper is positioned at a point between the selected UE and this BS, based on a weighted combination of their positions:
\[{Pos}_E = \alpha\times{Pos}_{UE_i} + (1 - \alpha) \times{Pos}_{BS} \tag{2}\]
where \(\alpha\) controls the proximity of the eavesdropper to the UE and BS, with lower values of \(\alpha\) placing the eavesdropper closer to the BS. Eavesdroppers need to support higher transmit power so they can overcome the attenuation, achieve the required signal-to-noise-plus-interference-ratio (SNIR) and capture signals. For this reason, \( P_{E_j}\), is randomly selected as:
\[P_{E_j} = \gamma*P_{UE_i}\  dBm \tag{3}\]

\subsection{Problem Formulation}
The next step is the creation of communication channels according to the aforementioned standards. We will focus on the first phase where Sounding Reference Signals (SRS) are created. These are signals that are sent by UEs to each BS and are used to estimate channel parameters (propagation paths, SE, losses, etc.). SRSs are transmitted through OFDM technology, pass through the channel and are subject to filtering and addition of white noise (AWGN). From these signals, each BS obtains the mandatory CSI, in order to differentiate legitimate users from possible attackers. One ML-enabled and three DL-enabled methods are proposed to solve this problem. All methods are examined either under known CSI or under additional location and power information of the UEs. These aspects are discussed in Subsection C.

The system is evaluated based on Secrecy Rate (SR), a metric that represents how much information can be securely transmitted without being intercepted by eavesdroppers. The secrecy rate for each channel, served by a BS \(b_w \in S_{BS}\) can be expressed as:
\[SR_{l_i} = [ C_{l_i, b_w} - C_{e_j, b_w}]^+ \tag{4}\]
where \(C_{l_i, b_w}\) is the channel capacity between the legitimate user \(l_i \in S_{LU}\) and the BS \(b_w\), \( C_{e_j, b_w}\) is the channel capacity between the eavesdropper \(e_j \in S_E\) and the same base station and \([.]^+\) denotes the positive part function, ensuring SR is non-negative. Throughout this paper, we will focus on the average secrecy rate which provides a measure of the overall secure communication efficiency in the system.
\[
\overline{SR} = \frac{1}{N} \sum_{i=1}^N SR_{l_i} \tag{5}
\]
This approach ensures that the system is optimized not only to maximize throughput, but also to prioritize secure communication between all legitimate users, minimizing the potential information gain of eavesdroppers. 

\section{ML/DL Based Eavesdropper Detection}

\subsection{Dataset Construction}
As stated in \cite{b1}, an essential step in developing ML and DL models is the training process. It is crucial to work with datasets that are precise, up-to-date, and have been evaluated. In this study, the dataset is generated using a MATLAB-based simulator, which models an Industrial Internet of Things (IIoT) scenario. The implemented simulation is modeled based on the 3GPP specifications TR 38.901 \cite{b14} and TR 38.843 \cite{b15} and has been properly adjusted to the characteristics of a B5G network which operates at Frequency Range 2. Channel parameters such as path loss, spectral efficiency, small- and large-scale fading are also taken into consideration, providing a detailed representation of the communication environment. 

In more detail, the simulator generates the communication channel between each pair of BS-UE, integrating factors such as carrier modulation, sampling rate, and the positions of UEs and BSs. Afterward, the channel matrices \(H\) for each UE-BS link are transformed into CSI images, with dimensions based on the number of subcarriers, OFDM symbols, ports, and BSs:
\[Image Dimensions=\]\[(NumSubCarriers \times NumOFDMSymbols \times\]\[NumPorts\times NumBaseStations) \tag{6}\]
These images capture the environment's multipath effects and signal attenuation under various propagation conditions, UE behaviors, and security threats. The data is labeled according to whether the UE is a legitimate user (0) or an eavesdropper (1). The final dataset consists of CSI images along with additional data about UE’s positions and transmission power. The generated dataset serves as the foundation for training advanced ML/DL models.

\subsection{ML/DL Models}
The first proposed ML model is a simple Random Forest (RF) classifier. The RF model constructs multiple decision trees, where each tree classifies the data based on splitting rules. At each node, the model assesses the purity (measured by the Gini index), which indicates how well the data is separated at that point. The model is trained using cross-validation on five different subsets of the training data. This method ensures robust model evaluation and at the same time prevents overfitting.

The second proposed model is a Deep Convolutional Neural Network (DCNN). Such networks are specifically designed for image analysis, making them an ideal choice for processing CSI images. The model was built using the Keras Sequential API and trained using the \textit{binary crossentropy} loss function. Below is its detailed structure:
\begin{itemize}
    \item Conv2D Layer: applies filters over the input data to extract important features while reducing dimensionality.
    \item Batch Normalization: stabilizes and accelerates training by normalizing activation values.
    \item ReLU Activation: introduces non-linearity to help the model learn more complex patterns.
    \item MaxPooling2D: reduces data dimensionality further
    \item Repeated Layers: two repetitions of Conv2D, Batch Normalization, and ReLU, aiming to refine feature extraction.
    \item GlobalAveragePooling2D: calculates the average of the features across each channel, reducing the 4D data to a 2D matrix.
    \item Dense Layer: a fully connected layer with a single neuron and a sigmoid activation to produce the binary classification output.
\end{itemize}
It is noted that this model processes only CSI images and labels. 

\begin{figure}[htbp]
\centerline{\includegraphics[width=1\linewidth]{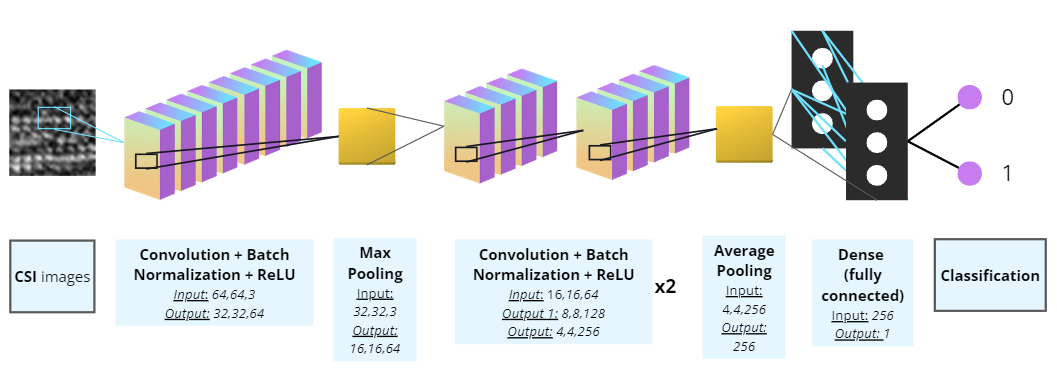}}
\caption{Proposed DCNN's Structure}
\label{fig2}
\end{figure}

The third model, DCNN 2, builds upon the previous architecture by integrating both visual data and categorical information, such as UE position and transmission power. The input's dimensions are \( (64, 64, 3 + 3) \), where the first three channels correspond to the color channels of the image, and the other represent the additional data. To manage the combined input, the model uses a Lambda layer to separate the non-image data. The image data is passed through the Convolutional and Pooling layers, while the position and power data are concatenated with the features extracted after the Global Average Pooling layer. Compared to the first DCNN, this model introduces two additional fully connected Dense layers with ReLU activation before the final \textit{sigmoid} layer. These extra layers enable the network to model more complex interactions between the image and non-image data.

The fourth and final model is a Long Short-Term Memory (LSTM) network, a type of Recurrent Neural Network (RNN) designed to process sequential data. Like the DCNN 2 model, it processes combined data inputs, but it introduces a temporal dimension, making it particularly suited for analyzing patterns over sequences. Before feeding the data into the LSTM network, pre-processing is required to reshape the input. The CSI images are flattened into sequences, forming an input shape of \((\text{batch\_size}, \text{time\_steps}, \text{features})\). Similarly, the position and power data are replicated across time steps, forming a sequence that matches the image sequence length. The exact levels are as follows:
\begin{itemize}
    \item LSTM Layer: processes the image data and encoding their information into a sequence of feature vectors.
    \item Dense Layer: a fully connected layer with ReLU activation extracts key features from the processed images.
    \item Dropout Layer: prevents overfitting by randomly dropping units during training.
    \item LSTM Layer: a second LSTM layer that outputs a final feature vector rather than a sequence.
    \item Repeated Layers: the position and power data are processed through a similar sequence of LSTM, Dense, and Dropout layers, treating them as time series data.
    \item Concatenate Layer: combines the outputs from both the image processing path and the position-power path.
    \item Dense Layer: a fully connected layer with ReLU activation processes the combined feature vector, allowing the model to learn more complex interactions.
    \item Output Dense Layer: uses softmax activation to produce probabilities for binary classification.
\end{itemize}

\begin{figure}[htbp]
\centerline{\includegraphics[width=1\linewidth]{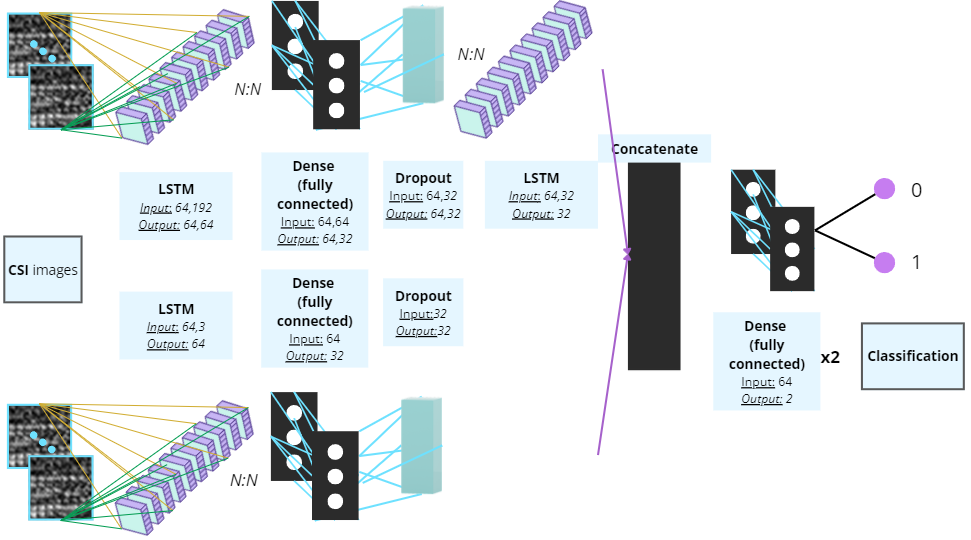}}
\caption{Proposed LSTM's Structure}
\label{fig3}
\end{figure}

\section{Performance Evaluation}

We consider a wireless B5G orientation, based on an IIoT scenario with dense UE placement. A cell-free 18 APs and 500 UEs topology is considered, where UEs are randomly distributed with a ratio of legitimate and malicious UEs of 65\%/35\%. The simulation includes both the transmission of SRSs from terminals to BSs (uplink) and the data transmission (downlink). Simulation parameters are summarized in Table I.
 
\begin{table}
    \centering
    \caption{Simulation Parameters}
    \label{tab:my_table}
    \begin{tabular}{|c|c|} \hline 
         \textbf{Parameter}& \textbf{Value/Assumption}\\ \hline 
         Frequency& 28 GHz (FR2)\\ \hline 
         Antennas' Height& BSs: 8m, UEs: 1.5m\\ \hline 
         UEs Mobility& 3 km/h\\ \hline 
         Noise& 5 dB\\ \hline 
         Transmit Power& BSs: 40 dB, LUs: 23 dB,
E: $>$23 dB\\ \hline 
         Subcarrier Spacing& 120 kHz\\ \hline 
 Bandwidth&400 MHz\\ \hline 
         Channel Modulation Symbols& 14\\ \hline 
         SRS Modulation Symbols& 12\\ \hline 
         Resource Blocks per Transmission& 60\\ \hline
         Number of Antennas& BSs: 32, UEs: 1\\ \hline
    \end{tabular}
\end{table}

Using the parameters presented above, the MATLAB B5G system and link level simulator produces the dataset that is used as input to the proposed ML and DL models. During the training phase of all approaches, a 80\%- 20\% training-test set split has been used. The problem of eavesdropper detection is examined as a classification one. Each model was hyper-parameter tuned via grid search, followed by cross-validation to ensure optimal performance. 

The models were then evaluated on a range of performance metrics to assess their effectiveness in accurately detecting eavesdroppers across a large set of UEs. One of the most commonly used evaluation metrics is Accuracy (Acc), which corresponds to the percentage of the total number of correct predictions divided by the total number of observations. However, relying solely on accuracy can be misleading in image classification, where anomalies indicating an eavesdropper occupy only a small portion of the data or maybe entirely absent. For this reason, additional metrics were employed, such as Sensitivity (Sen), also known as Recall, and F1-Score, which combines precision and recall to balance the evaluation between true positives and the potential impact of false positives and negatives.

Table II summarizes all performance metrics of the proposed ML and DL models in the eavesdroppers detection problem applied in the test set.

\begin{table}
    \centering
    \caption{Comparative Perfomance Analysis for all models}
    \label{tab:my_table_2}
    \begin{tabular}{|c|c|c|c|c|} \hline 
         &  &  \textbf{Accuracy}&  \textbf{Recall}& \textbf{F1-Score}\\ \hline 
         \textbf{DCNN 1}&  Not eavesdropper&  0.88&  1.00& 0.93\\ \hline 
         &  Eavesdropper&  1.00&  0.81& 0.90\\ \hline 
         \textbf{DCNN 2}&  Not eavesdropper&  0.98&  1.00& 0.99\\ \hline 
         &  Eavesdropper&  1.00&  0.98& 0.99\\ \hline 
         \textbf{LSTM}&  Not eavesdropper&  0.81&  1.00& 0.90\\ \hline 
         &  Eavesdropper&  1.00&  0.70& 0.82\\ \hline 
         \textbf{RF}&  Not eavesdropper&  0.95&  1.00& 0.97\\ \hline 
         &  Eavesdropper&  1.00&  0.93& 0.96\\\hline
    \end{tabular}
\end{table}

In general, all the selected models achieve above average to excellent performance. An important piece of information extracted from the above table is that all models have unit recall in the not eavesdropper class, which means that they do not make any false negative predictions. The worst performance is attributed to LSTM, which shows a tendency to misclassify eavesdroppers as typical UEs. DCNN 1 is the next one in the ranking and while it seems to have similar difficulties, its slightly increased performance suggests that it could be a competitive choice for this task, assuming further optimization. In the other models the differences are relatively small with DCNN 2 leading the race as it achieves the highest accuracy and highest f1-score in both classes, while its small lead in eavesdropper recall gives it a more balanced performance. However, one can consider that both DCNN 2 and RF can be used equally for this binary classification, depending on the application and its requirements.

\begin{figure}[htbp]
\centerline{\includegraphics[width=0.8\linewidth]{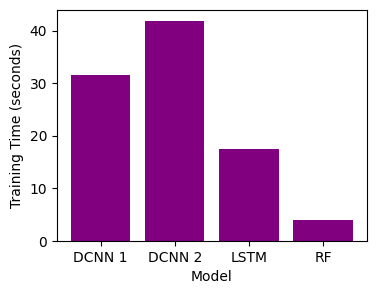}}
\caption{Training Time Comparison of Different Models}
\label{fig4}
\end{figure}

Many experiments emphasize the time it takes a model to train, as this is related to the computing power of the machines used. As shown below, all models take less than 50 seconds to train on the data. These times refer to a training data set for 360 UEs which corresponds to approximately 280 KB. ML models are clearly faster, with RF taking just 3 seconds. The complexity that characterizes DL models significantly increases their training time, with the highest for DCNN 2 reaching 48 seconds.

As mentioned before, the evaluation of the proposed PLS techniques will not be limited to the comparison of the above accuracy and complexity metrics. Delving into this problem, we performed another simulation of the system and the transmission process, in which we now entered as inputs the results of the two best-performing models. The purpose of this is to calculate the secrecy rates of the system, based on the formulas of Section II.

\begin{figure}[htbp]
\centerline{\includegraphics[width=0.8\linewidth]{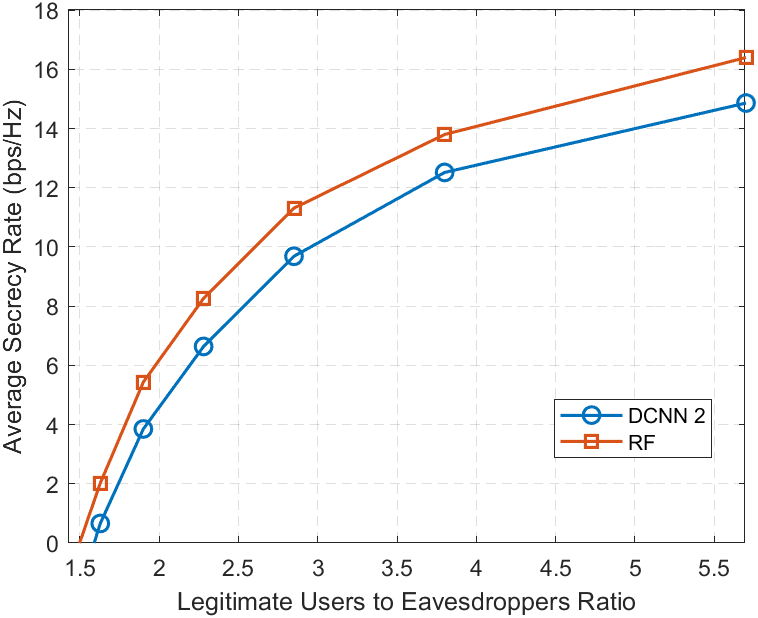}}
\caption{Average Secrecy Rate of system for DCNN 2 and RF}
\label{fig5}
\end{figure}

Figure 5 illustrates the average secrecy rate between the uplinks of the network as a function of the ratio of legitimate users (LUs) to eavesdroppers (Es). DCNN 2 (full version) and RF are compared.  It is evident that as the LU-to-E ratio increases, the average SR improves for both methods, because fewer transmissions and links with the BSs are subject to data interception. A point that needs attention is the fact that DCNN 2 demonstrates consistently lower secrecy rates than RF at all density ratios. This observation emphasizes the need to interpret the present graph in conjunction with the above results and not independently. In this context, the lower secrecy rates do not indicate that RF outperforms DCNN 2 in securing communications in this scenario. On the contrary, it is a more faithful representation of reality since it has correctly identified more eavesdroppers. Consequently, comparing the two models just proves that they do not show significant differences and can both serve this scenario. 

\section{Conclusions}

This paper investigates the use of ML/DL techniques to enhance PLS in B5G networks, focusing on identifying standard users and eavesdroppers within a heterogeneous network. Four models —DCNN 1, DCNN 2, LSTM, SVM, and RF— were evaluated, with RF excelling in performance and efficiency, and Convolutional Networks performing well on complex datasets. It should be noted that synthetic data contain limitations in terms of quantity and quality, because they may not always be available or representative of diverse network conditions. The absence of additional metrics, like SNR and throughput, insert additional limitations in PLS. However, the findings highlight the potential of AI to improve critical communications security, despite computational demands that may limit practical deployments in resource-constrained systems.

\end{document}